\documentstyle[twoside]{article}

\newcommand{\bfx}{{\bf x}}
\newcommand{\bfc}{{\bf c}}

\catcode`\@=11
\long\def\@makefntext#1{
\protect\noindent \hbox to 3.2pt {\hskip-.9pt  
$^{{\eightrm\@thefnmark}}$\hfil}#1\hfill}		

\def\thefootnote{\fnsymbol{footnote}}
\def\@makefnmark{\hbox to 0pt{$^{\@thefnmark}$\hss}}	
	
\def\ps@myheadings{\let\@mkboth\@gobbletwo
\def\@oddhead{\hbox{}
\rightmark\hfil\eightrm\thepage}   
\def\@oddfoot{}\def\@evenhead{\eightrm\thepage\hfil
\leftmark\hbox{}}\def\@evenfoot{}
\def\sectionmark##1{}\def\subsectionmark##1{}}

\input{ijmpc1.sty}
\input{psfig}

\begin{document}

\runninghead{Entropy and correlations in lattice gas automata without 
detailed balance} {Entropy and correlations in lattice gas automata without 
detailed balance}

\normalsize\textlineskip
\thispagestyle{empty}
\setcounter{page}{1}

\copyrightheading{}			

\vspace*{0.88truein}

\fpage{1}
\centerline{\bf ENTROPY AND CORRELATIONS IN LATTICE GAS AUTOMATA}
\vspace*{0.035truein}
\centerline{\bf WITHOUT DETAILED BALANCE}
\vspace*{0.37truein}
\centerline{\footnotesize OLIVIER TRIBEL\footnote{email address: 
{\tt Olivier.Tribel@ulb.ac.be}}}
\vspace*{10pt}
\centerline{\normalsize and}
\vspace*{10pt}
\centerline{\footnotesize JEAN PIERRE BOON\footnote{email address: 
{\tt jpboon@ulb.ac.be}}}
\vspace*{0.015truein}
\centerline{\footnotesize\it Center for Nonlinear Phenomena and Complex Systems}
\baselineskip=10pt
\centerline{\footnotesize\it Universit\'e Libre de Bruxelles, Campus Plaine, 
C.P. 231}
\baselineskip=10pt
\centerline{\footnotesize\it 1050 Brussels, Belgium}
\vspace*{0.225truein}
\publisher{(received date)}{(revised date)}
\textheight=7.8truein
\setcounter{footnote}{0}
\renewcommand{\thefootnote}{\alph{footnote}}

\vspace*{0.21truein}
\abstracts{We consider lattice gas automata where the lack of semi-detailed
balance results from node occupation redistribution ruled by distant 
configurations; such models with nonlocal interactions are interesting 
because they exhibit non-ideal gas properties and can undergo 
phase transitions. For this class of automata, mean-field
theory provides a correct evaluation of properties such as 
compressibility and viscosity (away from the phase transition), 
despite the fact that no $H$-theorem strictly holds.
We introduce the notion of {\em locality} -- necessary to 
define quantities accessible to measurements -- by treating the 
coupling between nonlocal bits as a perturbation. Then if we  
define operationally ``local'' states of the automaton -- whether 
the system is in a homogeneous or in an inhomogeneous state -- 
we can compute an estimator of the entropy and measure the local 
channel occupation correlations. These considerations are 
applied to a simple model with nonlocal interactions.}{}{}

\vspace*{10pt}

\vspace*{1pt}\textlineskip	
\section{Introduction}		
\vspace*{-0.5pt}
\noindent
Various lattice gas automaton (LGA) models lacking semi-detailed balance
(SDB) have been proposed because they exhibit interesting properties
for investigating phenomena which are not accessible to LGAs satisfying 
SDB.
LGAs violating SDB locally have been constructed for 
their operational value as possible candidates for the simulation of
high Reynolds number 
hydrodynamics\cite{henon:viscosity,dubrulle:viscosity,rivet:brisure}. 
Such systems
were subsequently reconsidered as a paradigmatic formulation of driven
systems\cite{ernst:nonBSD}.
The implications and consequences of the lack
of semi-detailed balance in lattice gas automata
are non-trivial,
in particular for what concerns their stationary distribution 
function and the very existence of an equilibrium distribution for such
systems which exhibit thermodynamic properties and transport properties 
correctly described by mean-field theory. This somewhat paradoxical situation 
has been the subject of consideration in the recent 
literature\cite{ernst:nonBSD,chen:generalized} where the question was raised 
whether the asymptotic state of systems without SDB can be qualified as an 
equilibrium state.
Certain LGAs without detailed balance were shown to 
approach a stable uniform state referred to as a non-Gibbsian 
equilibrium\cite{ernst:nonBSD} with nonlocal spatial
correlations, in contrast to the classical Gibbsian state of thermal 
equilibrium.

The problem has not been examined in depth for the case of LGAs where the 
lack of SDB results from node occupation redistribution ruled by
distant configurations; such models with nonlocal
interactions (NLIs) are interesting because they exhibit non-ideal gas
properties\cite{ot:interactions} and can undergo phase 
transitions\cite{appert:transition}. 
For this class of LGAs, mean-field
theory is also seen to provide a correct evaluation of properties such
as compressibility and viscosity\cite{ot:interactions,GEF}
(away from the phase transition) and static correlation functions 
are in accordance with
those observed in real fluids\cite{ot:interactions}. It has even been suggested
that a generalized version of the SDB condition could be applicable to recast 
LGAs with NLIs such that a generalized $H$-theorem would 
hold\cite{chen:generalized}. The situation calls for clarification, starting 
at the level of the basic statistical mechanical description of LGAs.

\section{Statistical Mechanics}\label{sec:statmech}
\noindent
We consider a set ${\cal S}$ of Boolean variables (bits), 
${\cal S}=\{s_{(i)},i=1,\ldots,bV\}$ with $s_{(i)}\in\{0,1\}$ (here $bV$ 
is simply an integer; $b$ and $V$ will be specified subsequently), 
and an update rule: ${\cal T}_\lambda:{\cal S}\rightarrow{\cal S}$, 
with $\lambda$ a set of parameters, allowing for example ${\cal T}$
to be drawn probabilistically among a set of rules.
This update rule is set according to certain constraints
which are chosen on the basis of physical requirements. 
${\cal S}$ and ${\cal T}$ define the lattice gas automaton.
The constraints on ${\cal T}$ are of two types:
\begin{itemize}
\item {\em Geometric constraints} lead to
 a spatial representation of the LGA, and impose certain symmetries
on this representation. This essentially amounts to defining a
set of relations $\{\left(i,\varphi(i)\right), i=1,\ldots,bV\}$ 
between indices of elements of~${\cal S}$. With these relations, 
the update rule ${\cal T}_\lambda$ may be
factorized into several steps, one of which is simply the copying of the
value of bit $i$ onto bit $\varphi(i)$; this step is called 
{\em propagation}
and the corresponding operator\fnm{a}\fnt{a}{This operator may itself be 
stochastic\cite{ot:levy}, but here we restrict ourselves to
 the deterministic case.}~
 is denoted by ${\cal P}$. 
An important feature of propagation is that it is a mere correspondence between
indices of elements of ${\cal S}$, independently of the values taken by 
these elements. 
To each $\varphi(i)$ we associate a geometric vector $\bfc_{(i)}$ 
in a suitable space, and these vectors (which may include the null vector)
are grouped into $b$ equivalence classes.
The vectors representative of these classes will be denoted by $\bfc_i$ 
(with $i=1,\ldots,b$); they must satisfy symmetry constraints in order 
to obtain a consistent geometric representation.
\item {\em Conservation laws} provide physical 
content. For example, denote by $n_{(i)}$ the value taken by the variable
$s_{(i)}$; then 
define a quantity $N[{\cal S}]=\sum_{i=1}^{bV}n_{(i)}$, which is identified as
the {\em number of particles}, and demand that 
$N[{\cal T}_\lambda{\cal S}]=N[{\cal S}]$. 
A similar conservation is generally required for {\em linear momentum}
${\bf p}=\sum_{i=1}^{bV}n_{(i)}\bfc_i$. 
\end{itemize}

We factorize the transformation ${\cal T}_\lambda$ as:
\begin{equation}\label{eq:locality}
{\cal T}_\lambda={\cal P}\circ {\cal U}_\lambda
\end{equation}
where ${\cal P}$ is the propagation as defined above. Now, 
through the operator ${\cal U}_\lambda$, it may happen that 
one can define {\em locality} .
If ${\cal U}_\lambda$ is such that there exists a decomposition of
${\cal S}$ into subsets $s._{(x)}$ in such a way that the restriction 
of ${\cal U}_\lambda$ over $s._{(x)}$,
denoted by ${\cal U}_\lambda|_{s._{(x)}}$, is completely defined and 
is an endomorphism:
\[
{\cal U}_\lambda|_{s._{(x)}}:s._{(x)}\rightarrow s._{(x)}\,,\; \forall (x)\,,
\]
--- i.e. if the values taken by the bits of $s._{(x)}$ 
after application of ${\cal U}_\lambda$ are entirely defined by 
their values before this application --- then  
the set of indices that define the subset $s._{(x)}$ is called a
{\em node}; the values taken by the bits of $s._{(x)}$ define the {\em
state of the node}. Given the above definition of $b$, $V$ is now seen
to be the number of nodes of the automaton universe and $b$ the number 
of channels per node. Then one may construct a representation of
the bits that belong to a given $s._{(x)}$
as `particles' and `holes' spatially located on the spot defined by $x$. 
The action of ${\cal U}$ is then {\em local} and ${\cal U}$ is 
called a {\em collision} operator (and will be denoted by ${\cal C}$). 
We call {\em L-subsets} the subsets $s._{(x)}$ thus defined.
Note that it is not mandatory that a LGA have such a property;
for example the models defined in section 3
have an update operator that cannot be decomposed strictly into propagation
and collision operators. 

\subsection{A Liouville $H$-theorem}\label{sec:liouville}
\noindent
We denote by $\Gamma$ the set of all possible sets ${\cal S}$.
We associate to each universe-state ${\cal S}$ a probability $P({\cal S})$
in a Gibbs ensemble and define a transition probability
${\cal A_{SS'}}$ between universe-states, such that the
system obeys a Chapman-Kolmogorov equation:
\begin{equation}
P(S')=\sum_{{\cal S}\in\Gamma}{\cal A_{SS'}}P({\cal S})\,,\; (S'\in\Gamma)\,.
\end{equation}
The following hypothesis on ${\cal A}$, the {\em semi-detailed balance} (SDB) 
condition:
\begin{equation}\label{eq:SDB}
\sum_{{\cal S}\in\Gamma}{\cal A_{SS'}}=1,
\end{equation}
suffices to prove an $H$-theorem\cite{henon:H};
then the {\em global entropy} 
\begin{equation}\label{eq:H}
H=-\sum_{{\cal S}\in\Gamma} P({\cal S})\ln P({\cal S})
\end{equation}
does not decrease under the action of ${\cal U}$.
Furthermore, as noted above, the propagation operator produces
a deterministic redistribution of the bits without introducing or 
removing information; consequently ${\cal P}$ does not modify the
value of $H$. 

For all practical implementations, $|\Gamma|$ is finite, but so 
large that no meaningful sampling is realizable; as a result the quantities
$P({\cal S})$ and ${\cal A_{SS'}}$ are not accessible to significant
measurements.

\subsection{Locality and nonlocality}\label{sec:locnonloc}
\noindent
If the updating operator ${\cal T}$ is such that it admits a decomposition
into a propagation operator ${\cal P}$ and a local collision operator
${\cal C}$, the local subsets $s_{(x)}$ define subspaces 
$\gamma_{(x)}$ of $\Gamma$, with $\Gamma=\bigotimes_{(x)} \gamma_{(x)}$.
If we assume that the $\gamma_{(x)}$'s are isomorphic,
then there exist local states $s$ such that the description of the automaton
in terms of the $s$'s is equivalent to its description in terms
of ${\cal S}$. Those states (of the nodes) 
are independent of each other for the local operator, so that we can define
a local probability $p(s_{(x)})$ in a (smaller) Gibbs ensemble. If the local
dynamics is described with a transition probability matrix 
$A_{ss'}$, then under the condition $\sum_{s\in\gamma}A_{ss'}=1 (s'\in\gamma)$, 
a {\em local $H$-theorem} holds\cite{henon:H} and the quantity
\begin{equation}\label{eq:h}
h(x)=-\sum_{s\in\gamma}p[s_{(x)} = s]\ln p[s_{(x)} = s]
\equiv -\sum_{s\in\gamma}p(s_{(x)})\ln p(s_{(x)})
\end{equation}
does not decrease under the action of ${\cal C}$.

In this local description, the action of ${\cal P}$ is unimportant
as it just provides at each iteration a fresh configuration of the bits of 
$s_{(x)}$ with probability 
$p(s)$ by definition; this probability is such
that $h$ does not decrease. The action of ${\cal P}$ is
of importance when one considers the connection between the local entropy 
(\ref{eq:h}) and
the global entropy~(\ref{eq:H}). If the propagation does not
produce correlations between the nodes, then 
the global entropy $H$ is entirely determined by the local entropy $h$:
\begin{equation}
H=\sum_{x}h(x).
\end{equation}

The existence of a notion of {\em locality} provides us with objects 
(the $s_{(x)}$'s)
which are accessible to measurements; then we can define the
local number of particles $n(x)$ and the local distribution function 
$\rho(x)$ by averaging $n(x)$ over the appropriate Gibbs ensemble:
$\rho(x)\equiv\langle n(x)\rangle$. 
We may choose the $s$'s as fundamental dynamical objects to which 
``physical'' properties are then associated.
In usual LGAs, this is accomplished by imposing constraints such as ``mass'' 
and ``momentum'' conservation. It is only because the update rule
allows a definition of locality --- i.e. the universe is a set of
independent subsets --- that such constraints have physical
justification since isolated  objects can now be considered. 
The local constraints
are chosen according to the problem to be investigated
and provide physical significance to the LGA model.

An important point is that the decomposition may be further pursued: 
we can describe the dynamics in terms of local 
space-velocity occupation $n_i(x)$ -- i.e. the value taken by the bit
$s_{(x)i}$, with $x=1,...,V$, and $i=1,...,b$ -- and of local single-particle 
space-velocity distribution functions 
$p[s_{(x)i}=1]=f_i(x) \equiv f(\bfx,\bfc_i)$. 
At this level of description an entropy is defined by:
\begin{equation}\label{eq:h_f}
h_f(x)=-\sum_{i=1}^{b}[f_i(x)\ln f_i(x)+\overline{f_i(x)}\ln\overline{f_i(x)}],
\end{equation}
(where $\overline{f_i(x)}\equiv 1-f_i(x)$)
which, under the Boltzmann {\em ansatz}, is simply equal to $h(x)$ (\ref{eq:h}).
The second term on the {\em r.h.s} of (\ref{eq:h_f}) stems from the  
correlations between 
``particles''(1-bits) and ``holes'' (0-bits) because of the Boolean nature of 
the variables. Omitting this term amounts to neglecting the important
contribution of the ``particle - hole'' correlations to the entropy.

If the operator ${\cal U}$ does not decompose ${\cal S}$ into independent,
disjoint subsets, then we have no consistent definition of locality. In 
particular, we cannot meaningfully define states $s$ and probabilities
$A_{ss'}$, and we have no definition of 
objects to which Statistical Mechanics can be applied: 
the semi-detailed balance condition is void, and no ``local''
$H$-theorem can exist.  An example is the class of automata
with ``nonlocal interactions'' (see section 3). 
We will however
argue that some models have update rules that allow for a weaker definition
of locality, i.e. there may exist a collection of disjoint 
subsets of ${\cal S}$, the bits of which are strongly coupled through
${\cal U}$, but only weakly coupled to the bits of other 
subsets. We will call these subsets {\em $\ell$-subsets}.
Then we can operationally define these
subsets as objects of Statistical Mechanics, even if they do not 
contain the full dynamics of the automaton. This is equivalent to 
ignoring all dependencies and correlations between ``nodes''. If the coupling
between ``nonlocal'' bits can be treated as a perturbation,
then we may be able to extract significant information on the dynamics 
{\em via} ``local'' quantities. However some basic elements necessary to 
establish the validity of mean-field theory (mainly the $H$-theorem) 
are then absent, and care must be taken in defining and using
``standard'' quantities.

\vspace*{1pt}\textlineskip	
\section{A Simple Model with ``Nonlocal Interactions''}\label{sec:model}		
\vspace*{-0.5pt}
\noindent
An example of a system where semi-detailed balance does
not exist is the LGA with ``nonlocal interactions'' introduced by Appert
and Zaleski\cite{appert:transition}, subsequently analyzed by Gerits
{\em et al.}\cite{GEF}, and generalized by Tribel and 
Boon\cite{ot:interactions}. Here we consider an utmost simplified
version of the automaton which nevertheless exhibits all essential
features and physical properties of the original model.

The LGA is composed of a set of bits (typically 393216) and an update rule
${\cal T}$ as described in Section 2. This rule admits
a decomposition into two operators; one of them is the
propagation operator ${\cal P}$ which copies the value contained
in $s_{(i)}$ onto the bit labeled $\varphi(i)$. The actual computation
of $\varphi(i)$ is performed by a computer routine which
is described elsewhere\cite{noullez:these}. The computation procedure
exhibits the following features: 
it defines the topology of the space used for the representation
of the state of the automaton on the surface of a torus and it yields 
six classes of equivalence of the $(i, \varphi(i))$'s to which we associate 
six vectors ${\bf c}_i$; for symmetry reasons these
vectors are chosen to be coplanar and normalized. 
The spatial representation of the lattice therefore has a toroidal topology 
and a planar geometry, the geometry of a triangular lattice with hexagonal 
symmetry. 

The second operator ${\cal U}$ which together with ${\cal P}$, 
composes the full updating operator is itself decomposed into two parts:
\[
{\cal U}={\cal U}_2\circ{\cal U}_1.
\]
${\cal U}_2$ groups the bits of ${\cal S}$ by subsets of six, and
re-shuffles the bits within the subsets. The grouping is such that, 
for any $i$ in a given $\ell$-subset $s_{(x)}$,
$\varphi(i)$ is in another $\ell$-subset $s_{(x)}$; for the re-shuffling, 
we use the rules of the FHP-I automaton (the collision outcome is governed 
by a Boolean parameter)\cite{frisch:hydrodynamics}. 
We denote by $s_{(x,j)}$ the
$j$-th bit of the subset $s_{(x)}$; since there is a one-to-one
correspondence between the indices $(i)$ and the indices $(x,j)$, we
may denote by $\varphi(x,j)$ the index of the bit where ${\cal P}$
copies the bit $(x,j)$, and by $\varphi^k(x,j)$ the index where
$(x,j)$ should be copied by applying $k$ times the operator ${\cal P}$. 
Finally, we will be interested in the $\ell$-subset to which 
$\varphi^k(x,j)$ belongs; its index will be $\chi(x,j,k)$.
 
The operator ${\cal U}_1$ re-shuffles the bits that belong to different 
$\ell$-subsets, according to the following procedure:
\begin{itemize}
\item draw a random number $D$, with equal probabilities among $\{1,2,3\}$ 
(corresponding to the three axial directions of the lattice); 
\item for each subset $s_{(x)}$, draw a random number $r$ according to 
a given probability distribution (which may be degenerate);
then, write the value $n_{(x,j)}$ taken by the bit $s_{(x,j)}$;
\item if: $s_{(x,D+1)}=1$, $s_{(x,D+2)}=0$, $s_{(\chi(x,D,r),D+1)}=0$,
and $s_{(\chi(x,D,r),D+2)}=1$, then exchange these values such that
$0$ becomes $1$ and vice-versa;
\item repeat with $D+4$ and $D+5$;
\item repeat the last two steps with $D$ replaced by $D+3$.
\end{itemize}
All the values of $D + E (E = 1,...,5)$ are mapped onto $(1,...,6)$.
The set of parameters $\lambda$ in $\cal U_\lambda$ is formed by the 
Boolean parameter of the collision, the direction $D$, and 
the distance $r$ (the set of distances). 

This model exhibits a phase transition when $r$ exceeds a given value 
(or an average value computed over a given probability distribution); 
away from the transition, the macroscopic properties of the LGA are
characterized by well-defined coefficients\cite{ot:interactions,GEF}.

\begin{figure}[htb]
\centerline{\psfig{figure=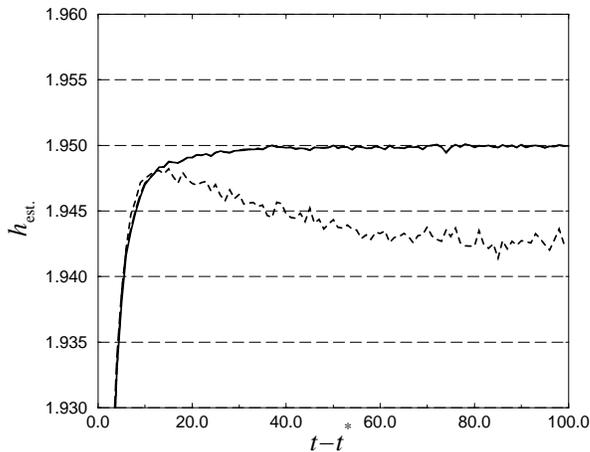,height=7cm}}
\caption{Estimated entropy, Eq.(8), as a function of time ($t > t^*$, see text). 
FHP automaton with strictly local collision rules and SDB (solid line), 
and LGA with NLI (interaction distance = 5 lattice units): 
(i) entropy estimated from the channel distributions $f_i$
(indistinguishable from full line), and (ii) entropy estimated from the 
configuration probability $p(s,t)$ (dashed line). 
All cases: average density $f=0.1$, lattice size $256\times 256$ nodes,
$t^*=500$;
time unit$=$automaton time step.}  
\label{fig:hest}
\end{figure} 

\begin{figure}[htb]
\centerline{\psfig{figure=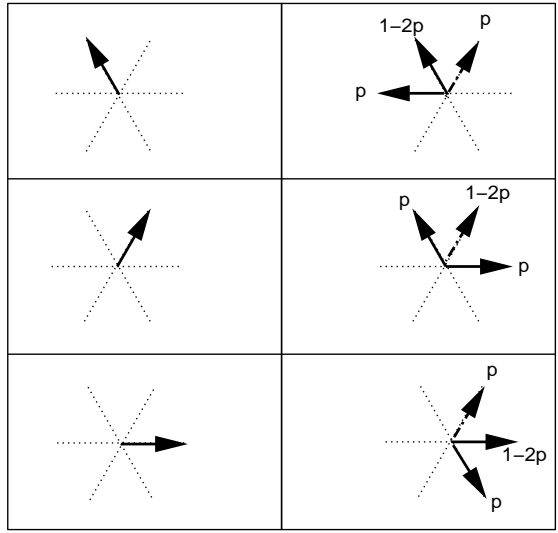,height=7cm}}
\caption{Example of transition probabilities yielding a seemingly valid SDB. 
Left boxes: initial configurations (arrows indicate particles).
Right boxes: output configurations (after non-local interaction and collision)  with the corresponding probabilities, assuming that the particles on the 
``pairing nodes'' are randomly distributed ($p$ is then simply one-third of 
the average density). Considering the sum of the transition probabilities
with an output configuration where the particle is in the upper right channel
(shown as dot-dashed arrow), the sum is clearly equal to one.}
\label{fig:example}
\end{figure}

\begin{figure}[htb]
\centerline{\psfig{figure=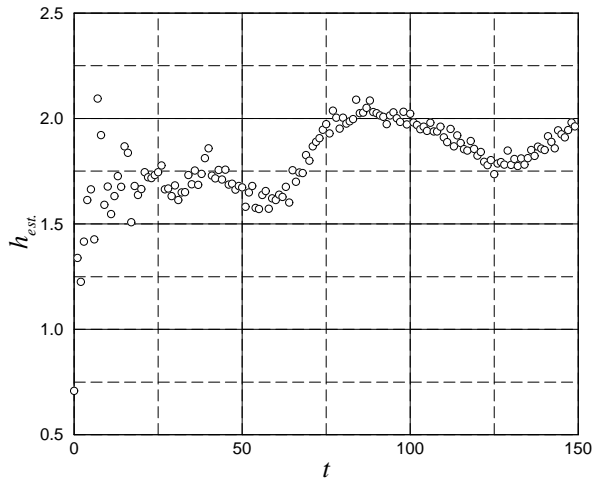,height=7cm}}
\caption{The entropy estimator, evaluated using the
occurrence frequency $\nu(s,t)$ of a ``local'' configuration at a given
node of the lattice, for 150 time-steps and 10000 realizations. The 
system is initialized with a step-function in the density field which 
is then relaxed; the estimator is measured on a node located at a 
position along the initial step. Parameters for the simulation: distribution 
of interaction distances between $1$ and $8$ lattice units, with
$p(r)\propto r^{-1}$; initial density: $f=0.1$ on one half of the lattice, 
$f=0$ on the other half; lattice size  $256\times 256$ nodes. No phase
transition occurs for these values of the parameters.}
\label{fig:h_grad}
\end{figure}

\begin{figure}[htb]
\centerline{\psfig{figure=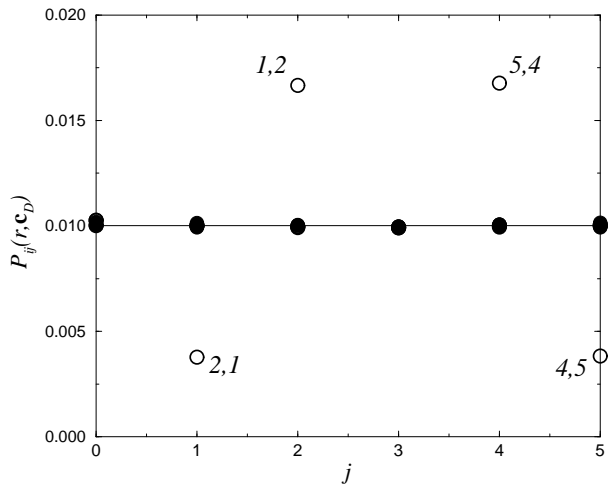,height=7cm}}
\caption{Nonlocal channel pair occupation probabilities, evaluated on 
nodes separated by interaction distance $r$ in LGA with NLI as described
in the text. 
The probabilities $P_{ij}(r,{\bfc}_D)$
are measured along a chosen direction $D$ (here $D=0$) when the interaction 
between channels labeled $i$ and $j$ occurs along that direction. 
The black dots are the superposition of data for $i=0\ldots 5$.
The density field is homogeneous; parameters of the 
simulation: $f=0.1$, $r=5$, lattice size $256\times 256$ nodes.}
\label{fig:nonlocalcorr}
\end{figure}

\begin{figure}[htb]
\centerline{\psfig{figure=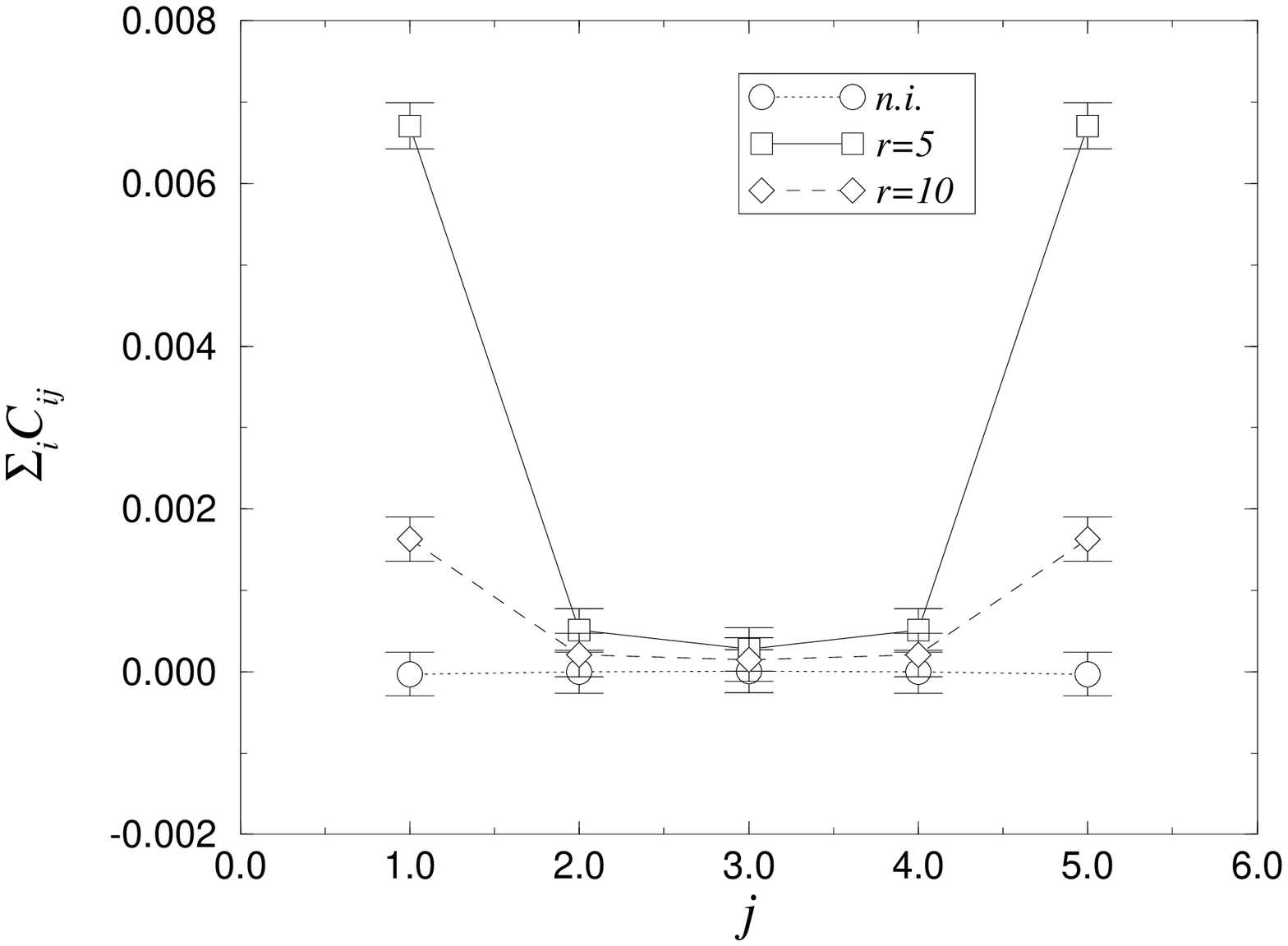,height=7cm}}
\caption{Local channel correlations 
($C_{ij}(0) = \langle n_i(\bfx)n_j(\bfx)\rangle-f^2$). 
Parameters of the simulation: $f=0.1$;  
lattice size $256\times 256$ nodes; interaction distance $r=5, 10$; for
comparison the open circles refer to the LGA without NLI. }
\label{fig:localcorr}
\end{figure}


\subsection{Entropy}
\noindent
We may operationally define ``local'' states $s(x,t)$ ($\ell$-subsets) of the 
automaton (at time $t$), and, if the LGA is in a homogeneous 
state, we can evaluate the occurrence probability $p(s,t)$ of a configuration 
$s$ by measuring the occurrence frequency $\nu(s,t)$ over the whole lattice at 
each time step. 
A second measure of $p(s,t)$, closer to its 
definition, is the occurrence frequency at a given ``location'' of the 
automaton, over a large number of realizations; at each run the automaton
is initialized independently  with a given set of macroscopic constraints. 
This (time-consuming) procedure is necessary if the automaton is not in 
a homogeneous state at all times. The occurrence probability changes 
with time, and so does the quantity:
\begin{equation}\label{eq:hest}
h_{\rm est.}(t) = -\sum_s \nu(s,t) \ln \nu(s,t).
\end{equation}
In an automaton with strictly local rules and satisfying SDB (\ref{eq:SDB}), 
this quantity will increase  monotonously towards a maximum value; 
in a nonlocal automaton, this increase is not guaranteed.

Consider a FHP type LGA subject to a constraint realized by imposing a 
systematically oriented output configuration for binary collisions; then
after some time $t^*$, the constraint is relaxed by restoring the usual 
rule with output configurations equally distributed along the three directions.
During the first phase the entropy decreases from its initial value, then
at time $t^*$ it starts increasing and levels off at its maximum value. 
This example is illustrated in Fig. \ref{fig:hest} for LGAs with and 
without NLIs. For a LGA with strictly local rules and satisfying SDB, we
observe monotonic increase of $h_{est}$ as expected; for the   
LGA with NLIs, the entropy estimated using the configuration probability
$p(s)$ shows non-monotonic increase and stabilizes at some value below the
plateau value obtained for the LGA without NLIs. 
Alternative interpretations are possible:
(i) if $h$ is a valid quantity and is correctly measured
--- i.e. $\nu$ is a good estimator of $p$ --- 
then we conclude that SDB is violated and the $H$-theorem does not hold;
(ii) the local states are here ill-defined, and we conclude
that $h$ is not a meaningful quantity and that the $H$-theorem at this level
of description is irrelevant.
Now LGAs without SDB can also be viewed as operational models to simulate 
non-equilibrium systems\cite{bussemaker:selforganization}; 
but LGAs with NLIs would be
driven systems without either boundaries or any external ``field'', 
which renders the ``non-equilibrium constraints'' rather unphysical. 

\subsection{Transition probability}
\noindent
Using the weak definition of locality (section 2.2), 
we define as above ``local'' states of the automaton. 
Then a transition frequency $F_{ss'}$ can also be defined and 
we can measure $F_{ss'}$ either by averaging over the whole ``lattice''
(if the state is homogeneous) or by averaging over many realizations 
(Gibbs ensemble). Again we must stress that this is not a strictly correct 
estimator of the transition probability $A_{ss'}$ which does not exist
in automata with nonlocal rules where the transition frequency depends 
on the configuration of the whole set ${\cal S}$ (i.e. on the density field 
over the whole lattice); the transition frequency
 measured in an inhomogeneous state 
differs from its value obtained in a homogeneous state.

Consider a LGA model with ``nonlocal'' interactions where we can measure the 
transition  {\em frequency} $F_{ss'}$. Here we are only interested in ``local'' 
configurations, but we can make no {\em a priori} assumption either on the 
direction of the interaction or on the state of the ``pairing'' nodes.
We then define a ``mean-field'' transition 
{\em probability} $A^{\rm m.f.}_{ss'}$, evaluated by 
considering an arbitrary node and assuming that the configurations of all 
the other nodes are given by a probability distribution 
$\tilde{p}(s)$.\fnm{b}\fnt{b}{$F_{ss'}$ is a statistical measure
of the unknown quantity $A_{ss'}$, while $A^{\rm m.f.}_{ss'}$ is
a theoretical evaluation of this quantity under the mean-field hypothesis.}
If we assume that the bits are all uncorrelated and that each channel has 
occupation probability $f$ (the average density per channel), any 
configuration has probability $\tilde{p}(s)=f^{N(s)}(1-f)^{[b-N(s)]}$, 
where $N(s)$ is the number of occupied channels in configuration $s$.
Then the interactions are equally probable along any direction, and
since the interaction rules have the same symmetries as the
underlying lattice, each momentum change can occur with the same probability 
as the reverse change. Therefore in this mean-field picture, the collision
rules are such that SDB is satisfied,\fnm{c}\fnt{c}{Here even detailed balance 
is satisfied.} and so is the complete update rule, i.e.
\begin{equation}
\sum_{s\in\gamma}A^{\rm m.f.}_{ss'}= 1.
\label{mfsdb}
\end{equation}
Note that (\ref{mfsdb}) does not mean that the matrix elements $A_{ss'}$ are 
identical in systems with and without nonlocal interaction; only
the sums over all initial states are equal to one in both cases; for an
example, see Fig. \ref{fig:example}. However the time behavior of
the entropy estimator is different in the two types of system; when NLIs are 
present the entropy estimator is not a monotonously growing function, see 
Fig.\ref{fig:hest} (homogeneous case) and Fig.\ref{fig:h_grad} 
(inhomogeneous case). 
Direct measurement of the transition frequencies reveals indeed that the 
mean-field approximation is very poor (we find significant deviations to
Eq.(\ref{mfsdb})) in accordance with the fact that the $H$-theorem does 
not hold. 
We conclude that the mean-field transition probability is not an appropriate 
quantity to correctly describe the dynamics of LGAs with nonlocal rules. 
Furthermore the reasoning does not include the interaction distance, so 
omitting one of the most essential features of the model.

\subsection{Boltzmann {\em ansatz}}
\noindent
In LGA theory, the Boltzmann {\em ansatz} allows to establish the explicit
connection between different levels of description of the $H$-theorem 
(hence the existence of a known equilibrium), and provides a way to obtain  
the explicit expression of the channel occupation distributions $f_i(\bfx,t)$. 
The first point is essentially of theoretical importance; the second point
has crucial consequences for the computation of the properties of LGAs.

Nonlocal interactions occur only between nodes with certain configurations, 
producing output configurations deterministically. If we denote the 
direction of interaction at time $t$ by $D$, then only  particles
on the pairs of channels $\{D+1,D+2\}$ and $\{D+4,D+5\}$ (modulo 6) can
interact, thereby exchanging their momentum. As a result, a configuration 
where pairs of channels ($\{D+i,D+i \pm 1\}$) are occupied by ``converging'' 
particles becomes more probable than a configuration with ``diverging'' 
particles. Since the direction of interaction is randomly chosen at each 
time step, this effect is not visible in single-node configurations, but 
must be expected to show up in nonlocal channel-pair correlations 
$C_{ij}(r,{\bfc}_D)$ defined by
\begin{eqnarray}\label{eq:defC}
P_{ij}(r,{\bfc}_D)&\equiv&\langle n_i({\bfc})n_j({\bfx}+r{\bfc}_D)\rangle
\nonumber\\
&\equiv&f^2+C_{ij}(r,{\bfc}_D),
\end{eqnarray}
with $f$ the average density.
The Boltzmann {\em ansatz} assumes that
\begin{equation}
C_{i,j}(r,{\bfc}_D)=f(1-f)\delta_{ij}\delta_{r0}.
\end{equation}
This is indeed the case in a LGA with SDB; however, in automata with
nonlocal interactions, or more generally without SDB, correlations
arise. For the present model, one expects to 
observe correlations between channels oriented along a 60-deg. angle, since 
only particles on channels with neighboring indices do interact.
Figure \ref{fig:nonlocalcorr} shows the correlations measured in a simulation;
the results are clearly in agreement with the predicted effect. The LGA Enskog 
formalism\cite{bussemaker:analysis} can be generalized to incorporate higher 
orders in the correlations; a comparative analysis between theory and 
simulation results will be presented elsewhere.

Nonlocal interactions are sources of correlations.
Once they are created, these correlations propagate, and, 
at sufficiently low density, they may survive until the involved particles 
reach the same node (following their initially converging trajectories).  
The results given in Fig. \ref{fig:localcorr} show indeed that {\em nonlocal} 
correlations (induced by NLIs) create {\em local} correlations (i.e. between
adjacent channels on the same node). The existence of such correlations 
explains the difference between the values of the entropy estimated {\em via} 
the single-particle distribution functions $f_i$, and of the entropy estimated 
{\em via} the configuration probability $p(s)$ (see Fig. \ref{fig:hest}): 
local correlations create an entropy deficit. Indeed it is a general 
property that the entropy computed from the full statistics of
a set of Boolean variables is lower than, or equal to the entropy
computed from the set of individual statistics\cite{BR}: 
correlations contain information, and therefore reduce the entropy.

\vspace*{1pt}\textlineskip	
\section{Comments}		
\vspace*{-0.5pt}
\noindent
We have presented a description of lattice gas automata using a rigorous 
definition of locality. The basic objects to be used in LGA statistical 
mechanics and in simulation measurements are defined in terms of the operator
which updates the bits of the automaton. The analysis introduces a distinction 
between (i) LGAs where a strict notion of locality exists, in which case
the semi-detailed balance condition is well defined (whether satisfied
or not), and (ii) LGAs with a weak notion or no notion of locality,
where the states of the node do not determine the full dynamics, and 
where semi-detailed balance loses meaning. For the first
class of models the question of the existence of a local equilibrium is 
quite relevant. In automata of the
second class, the existence of a global equilibrium is, in general, 
not related to the existence of a local equilibrium, and the Boltzmann 
hypothesis is invalid at all ranges.  
We have proposed a weaker definition of locality and of local states which 
can  be used in appropriate cases; with this lax definition,  mean-field 
analysis can be justified for the computation of thermodynamic and transport 
properties of the lattice gas.

\nonumsection{Acknowledgements}
\noindent
OT has benefited from a grant from the Fonds pour la Formation \`a la 
Recherche dans l'Industrie et l'Agriculture (FRIA, Belgium). JPB acknowledges 
support from the Fonds National de la Recherche Scientifique (FNRS, Belgium).

\nonumsection{References}

\end{document}